\documentclass[twocolumn,showpacs,preprintnumbers,amsmath,amssymb]{revtex4}
\usepackage{graphicx}
\usepackage{dcolumn}
\usepackage{bm}

\def\C{\mbox{\bf C}}
\def\M{\mbox{\bf M}}
\def\P{\mbox{\bf P}}

\def\ca{\mbox{Ca$^{2+}$}}

\def\na{\mbox{Na$^{+}$}}
\def\k{\mbox{K$^{+}$}}
\newcommand{\be}{\begin{equation}}
\newcommand{\bea}{\begin{eqnarray}}
\newcommand{\ee}{\end{equation}}
\newcommand{\eea}{\end{eqnarray}}
\begin{document}

\title{Synaptic plasticity with discrete state synapses} 

\author{Henry D.I. Abarbanel}
\affiliation{Department of Physics and Marine Physical Laboratory (Scripps
Institution of Oceanography)}
\altaffiliation[Also at]{Institute for Nonlinear Science} 
\altaffiliation{University of California, San Diego;
La Jolla, CA 92093-0402  USA}

\author{Sachin S. Talathi}
\affiliation{Department of Physics and Institute of Nonlinear Science}
\altaffiliation{University of California, San Diego;
La Jolla, CA 92093-0402  USA}
\email{talathi@physics.ucsd.edu}

\author{Leif Gibb}
\affiliation{Graduate Program in Computational Neurobiology, and Institute
for Nonlinear Science; Division of Biological Sciences}
\altaffiliation{University of California, San Diego;
La Jolla, CA 92093-0402  USA}

\author{Misha Rabinovich}
\affiliation{Institute for Nonlinear Science}
\altaffiliation{University of California, San Diego;
La Jolla, CA 92093-0402  USA}

\date{\today}

\clearpage

\begin{abstract}

Experimental observations on synaptic plasticity at individual
glutamatergic synapses from the CA3 Shaffer collateral pathway onto
CA1 pyramidal cells in the hippocampus suggest that the transitions
in synaptic strength occur among discrete levels at individual
synapses (~\cite{Peter}
and S. S.-H. Wang, unpublished data used with the authors'
permission). This happens for both long term potentiation (LTP) and
long term depression (LTD) induction protocols. O'Connor,
Wittenberg, and Wang have argued that three states would account
for their observations on individual synapses in the CA3-CA1
pathway. We develop a quantitative model of this three state system
with transitions among the states determined by a competition
between kinases and phosphatases shown by O'Connor et al. to be
determinant of LTP and LTD, respectively. Specific predictions for
various plasticity protocols are given by coupling this description
of discrete synaptic AMPA conductance changes to a model of
postsynaptic membrane potential and associated intracellular
calcium fluxes to yield the transition rates among the states. We
then present various LTP and LTD induction protocols to the model
system and report the resulting whole cell changes in AMPA
conductance. We also examine the effect of our discrete state
synaptic plasticity model on the synchronization of realistic
oscillating neurons. We show that one-to-one synchronization is
enhanced by the plasticity we discuss here and the presynaptic and
postsynaptic oscillations are in phase. Synaptic strength saturates
naturally in this model and does not require artificial upper or
lower cutoffs, in contrast to earlier models of plasticity.

\end{abstract}

\pacs{Valid PACS appear here}

\maketitle

\clearpage

\noindent
\section{Introduction}

Experiments on synaptic plasticity at individual synapses in CA3 to
CA1 hippocampal pathways reveal an ``all-or-none" change in their
synaptic strength (~\cite{Peter} and S. S.-H. Wang, unpublished data used with the
authors' permission). Indications of this were seen a decade ago
(C. F. Stevens and Y. Wang, personal communication). The recent
measurements have given substantial standing to the notion that
single synapses may operate as a discrete state system in their
plastic changes associated with long-term potentiation (LTP) and
long-term depression (LTD).

In this paper we first explore a general formulation of the
possibility that individual synapses can express a finite number of
discrete levels of conductance rather than the continuous or graded
or analog picture often formulated ~\cite{Castellini,Kar,Shouval,Abar1}. ~\cite{Peter} have
commented on the positive consequences for reliability of neural
memory from discrete state synapses. In addition there is
computational evidence that at an individual dendritic spine the
number of free \ca ions at the peak of a \ca transient evoked by
synaptic stimulation may be small ~\cite{Frank}. A
synapse which must respond to this very small signal might well
select a strategy of an ``all-or-none'' response in adjusting its
strength as a means of achieving a measure of reliability.

After a brief discussion of a general formulation of an $L$ level
synapse, we focus our attention on $L = 3$, which is suggested by
the recent data of O'Connor et al. (D. H. O'Connor, G. M.
Wittenberg, and S. S.-H. Wang, unpublished data used with the
authors' permission). With three levels we explore the transitions
among the levels using observations of O'Connor et al. With some
general arguments based on the measurements, we are able to
establish values for the normalized conductances of the individual
$L = 3$ levels, and we suggest an experiment which will determine
an interesting ratio among the transition rates.

The transition rates themselves depend on a model for how complex
chemical pathways lead to the change in AMPA conductance and the
number of AMPA receptors at any synapse ~\cite{Car,Lisman,Winder,Contract,Malin,Sheng}. We adopt
a version of our earlier model ~\cite{Abar1} of these
processes to provide a basis on which to make quantitative
predictions of the outcome of various LTP/LTD induction protocols
on the changes in AMPA conductance of a postsynaptic cell. We
explore spike time dependent protocols  as well as presentation of
spike bursts of various frequencies to the postsynaptic cell. These
compare well to experiment, in particular to results presented by
~\cite{Witten}, and predictions are made for various spike
timing experiments. We also explore the following setup: a
periodically firing conductance based presynaptic neuron provides
excitatory synaptic input to a periodically firing conductance
based postsynaptic neuron. In certain ranges of frequency and
conductance strength for the excitatory connection these neurons
synchronize ~\cite{Pikov}. We explore the effect on this
synchronization of our discrete synaptic strength model and show
that the regime of one-to-one synchronization is significantly
enlarged ~\cite{Zhig} for some conductance strengths and
that the synchronization is an in-phase firing of the two neurons.

It is an important feature of models built on a finite number of
discrete synaptic levels that the synaptic strength (AMPA
conductance) is always bounded above and below. This is in contrast
to models developed over the years, including our own ~\cite{Novot}, 
which do not have this property.

\section{Methods}

\subsection{Discrete State Synapses; Transition Rate Models}

The data of O'Connor et al. suggest that three discrete states of
AMPA conductance are found at individual synapses. There are
suggestions of other numbers of discrete levels (~\cite{Peter}; C. F. Stevens and Y. Wang, personal communication).
 If the
total number of levels is L and is indexed by $l = 0,1, 2, ...,
L-1$, and there are $N_S$ synapses indexed by $n = 1, 2, ..., N_S$,
then we represent the occupation of synapse n in state l at time t
by $N_{l}^{n}(t)$ . These occupation numbers are either zero or
unity. They can change in time due to LTP/LTD induction protocols
or other biological processes. The average occupation number in
state $l$ is given by \be p_l(t) = \frac{1}{N_S}\sum_{n=1}^{N_S}
N^{(n)}_l(t).
\ee

These $p_l(t)$ will constitute the main dynamical variables of our
model. They are taken to satisfy linear rate equations of the form
\be
\frac{d p_l(t)}{dt} = \sum_{k=0}^{L-1}  \biggl\{W_{(k\to l)}(t) p_k(t) - W_{(l \to k)}(t)
p_l(t) \biggr \}.
\ee
The transition rates $W_{(k\to l)}(t), W_{(l \to k)}(t)$ are
selected to assure
\be
\sum_{k=0}^{L-1}p_k(t) = 1.
\ee
As long as the number of active synapses $N_S$ is unchanged, this
last equation follows from the definition of $p_k(t)$.
In the limit of $N_S$ large, we assume that the $p_l(t)$ remain finite; this is
a standard assumption about such a limit for the description of a large number of
objects each having a discrete set of states.

One can
collect the average occupation numbers into an L-dimensional vector
$\P(t)=(p_0(t),p_1(t), ...p_{L-1}(t))$ which satisfies
\be
\frac{d \P(t)}{dt} = \M(t) \cdot \P(t),
\ee
and the conservation rule means that $\M(t)$ always has at least
one zero eigenvalue  ~\cite{Law,Van} the dynamics of
$\P(t)$ takes place in the $L-1$ dimensional space orthogonal to
the constant L-dimensional vector $\C = (1,1,1,...,1)$.

The effect of having this constraint may be seen in the
decomposition of $\P(t) = \C + \P_{\perp}(t)$ where $\P_{\perp}(t)
\cdot \C = 0$. The dynamics of motion for $\P_{\perp}(t)$ is
\be
\frac{d \P_{\perp}(t)}{dt} = \M(t) \cdot \P_{\perp}(t) + \M(t) \cdot \C,
\ee
which is driven motion of the vector $\P_{\perp}(t)$ spanning the
$L-1$ dimensional space orthogonal to $\C$. $\P_{\perp}(t)$ is
defined up to a rotation about $\C$. This dynamical description is
similar to that of driven precession of a spin in a time dependent
magnetic field.

Each discrete level $l = 0,1,...,L-1$ has an AMPA conductance $g_l$
normalized to baseline. The normalized, dimensionless AMPA
conductance of the neuron with $N_S$ synapses is
\bea
G_{AMPA}(t) &=& \sum_{n=1}^{N_S}\,\sum_{l=0}^{L-1}g_lN_l^{n}(t)
\nonumber \\ &=& N_S \sum_{l=0}^{L-1}g_lp_l(t).
\eea
This means that the quantity
\be
\frac{G_{AMPA}(t)}{N_S} = \sum_{l=0}^{L-1}g_lp_l(t),
\ee
the normalized, dimensionless AMPA conductance per synapse, is
independent of the number of synapses when $N_S$ is large,
depending only on the average occupation number of the synaptic
levels and the conductance associated with each level. By
definition, before any LTP/LTD induction protocols are presented to
the postsynaptic neuron, this quantity is equal to one:
\be
\frac{G_{AMPA}(t=0)}{N_S} = 1.
\ee
In our work below, we report the quantity $\frac{G_{AMPA}(t)}{N_S} - 1$ as the output
from our simulation of various induction protocols.

To fully specify the model of synaptic plasticity associated with
the presence of discrete levels, we must identify the conductances
$g_l$ of each level, and, through some form of dynamics of the
postsynaptic neuron, determine the transition rates. In the next
section we do this for the suggested three state model of O'Connor
et al. However, if observations indicate that there are $L
\ne 3$ levels operating at some synapses, then the general formulation presented
here will cover that situation as well.

\subsection{Three State Model}

If there are three states $l = 0, 1, 2$ then we need to identify
three discrete level conductances $g_0, g_1, g_2$ and the transition rates
among the levels. O'Connor et al. call their three levels ``low"
(state 0 here), high (state 1 here), and ``high locked-in" (state 2
here). They suggest that transitions associated with LTD protocols
connect state 1 to state 0, and transitions associated with higher
frequency protocols, typically leading to LTP, connect state 0 to
state 1 and state 1 to state 2. They also note that when an LTD
protocol is applied {\em following} a saturating LTP protocol to a
population of synapses, the synapses cannot be depressed as fully
as when the LTD protocol is applied to a naive population of
synapses. The amount of depotentiation possible decreases over the
10 minutes following LTP induction. This led them to suggest the
presence of a ``high locked-in" state, called H$^{*}$ by them; we
call this state 2. They do not require transitions between state 2
and state 0 to account for their data, and we assume none as well.

Using these observations we associate an ``LTD transition rate"
$g(t)$ with the transition between state 1 and state 0. Similarly
we associate an ``LTP transition rate" $f(t)$ with transitions
between state 0 and state 1. Loosely speaking we think of $g(t)$ as
an aggregated action of phosphatases leading to the
dephosphorylation or removal of synaptic AMPA receptors and $f(t)$
as the aggregated action of kinases operating in the opposite
fashion ~\cite{Lisman,Winder}. In the next
section we will specify how one evaluates these transition rates
from a dynamical model of the postsynaptic neuron, but for the
moment we note that $f$ and $g$ will depend on the elevation of
intracellular postsynaptic calcium concentrations above the
equilibrium level $C_0 \approx 100\,$ nM. Denoting the time course
of postsynaptic intracellular calcium concentration as $Ca(t)$, we
define
\be
\Delta Ca(t) = \frac{Ca(t) -C_0}{C_0},
\ee
and the transition rates $f(t),g(t)$ are determined by $\Delta Ca(t)$ in a manner
specified in the next section.

The transition between state 1 and the ``high locked-in" state
called 2 is taken to be proportional to the $0 \to 1$ transition
rate $f(t)$. If one had more detailed information on the
biophysical kinase and phosphatase pathways, one could replace this
simple assumption by a more complex quantity. We take this
transition rate as $b f(t)$ with $b$ a constant to be determined.

The picture outlined by O'Connor et al. does not suggest a
transition from state 2 to state 1, but we find it is necessary.
For the moment we call this transition rate $h(t)$, and we will
argue that it is proportional to $f(t)$. $h(t)$ cannot be zero, if
the transition rate framework is to be consistent with
observations.

This discussion leads us to the transition rate (or ``master'') equations
associated with the scheme depicted in Figure 1:
\bea
\frac{dp_0(t)}{dt} &=& -f(t)p_0(t) + g(t)p_1(t) \nonumber \\
\frac{dp_1(t)}{dt} &=& f(t)p_0(t) + h(t)p_2(t) - g(t)p_1(t) -
b f(t) p_1(t) \nonumber \\
\frac{dp_2(t)}{dt} &=& bf(t)p_1(t) - h(t)p_2(t).
\label{master}
\eea
By construction
\be
\frac{d (p_0(t) + p_1(t) + p_2(t))}{dt} = 0.
\ee

Under prolonged stimulation the postsynaptic intracellular calcium
levels reach approximately constant values, and we can ask what is
the behavior of $\P(t) = [p_0(t),p_1(t),p_2(t)]$ under such
circumstances. This means the functions $f(t)$ and $g(t)$ are
thought of now as constant in time with magnitude determined by the
saturated level of \ca. We associate this value of $\P$, after long
constant \ca elevation, with the fixed point of the equations
(\ref{master}) for $\P(t)$. If the induction protocol lasts a time
$T_I$, then the state long after $T_I$ starting with an initial state
$\P(0)$ will be the fixed point
\be
\P_{\mbox{fixed point}} = \frac{(gh,fh,b f^2)}{h(f+g) + b f^2}.
\label{fixed}
\ee

In our models, low values of saturated intracellular calcium
elevation $\Delta Ca$ are connected with LTD and higher values,
with a competition between LTD and LTP ~\cite{Yang}. The
specific form of the connection between $\Delta Ca$ and $f$ and $g$
will be given shortly, but their general dependence is shown in
Figure 2 ~\cite{Brad}. In an LTD protocol $g \ne 0$ but
$f
\approx 0$. In an LTP protocol both $f$ and $g$ may be nonzero.

If we take $h = a g$, then a saturating LTD protocol, with $g \ne 0, f \approx 0$, will
deplete both states 1 and 2, leading to a final state $\P = (1,0,0)$.

If, however, we apply a saturating LTP protocol where neither $f$
nor $g$ is zero, thus arriving at (\ref{fixed}), and \textit{then}
apply a saturating LTD protocol, the choice $h = a g$ will lead us
back to the state $\P = (1,0,0)$, which is not what is observed.
Indeed, O'Connor et al. note that the state reached by a saturating
LTP protocol depotentiates to a state intermediate between all
synapses in state 0 and the fully saturated state; namely, our
fixed point (\ref{fixed}).

If we choose $h = a f$, then this saturating LTD protocol following the saturating LTP protocol leads
us to
\be
\P = \frac{(a(f+g),0,b f)}{a(f+g) + b f},
\ee
namely we depopulate state 1 due to the action of $g$. This is the
kind of depotentiated, but not baseline, state seen by O'Connor et
al.

We conclude that the choice $h = a f$ is consistent with the
observations, and we cannot have $a = 0$. If $ a = 0$,  the high
locked state would be totally populated by a strong LTP protocol
and the synapse would not leave that state. Indeed, O'Connor et al.
indicate that after such a strong LTP protocol (two rounds of
theta-burst stimulation), most but not all synapses, about 80\%,
are in state 2.

This completes the general formulation of the three level
transition rate model. We now turn to the determination of the AMPA
conductances $g_l$ in each level $l = 0, 1, ..., L-1$ from the data
presented by O'Connor et al. Then we discuss a conductance based
neural model for the postsynaptic cell which permits us to
translate electrophysiological activity into transition rates
useful in the equations of $\P(t)$.

\subsection{Numerical Method}

All the simulations for the model presented in this work were
written in C and used a 4th order Runge-Kutta algorithm with a
fixed time step of 0.01 ms. They were run under Linux on a computer with an
Athlon 2400 MHz processor.

\section{Results}

We first establish, using the measurements of O'Connor et al., the
values of the normalized, dimensionless AMPA conductances of the
three levels at an individual synapse. Our arguments show that they are determined independently of the specific model for the transition rates.
 Then we develop a model for
the transition rates which will allow us to make predictions about
the response of the cells to various LTP/LTD protocols.

\subsection{Determination of the Discrete Level Conductances}

At $t=0$ the observed average occupation of levels is observed to be
 $\P(0) = (\frac{3}{4},\frac{1}{4},0)$. This means the normalized
AMPA conductance is
\be
\frac{G_{AMPA}(0)}{N_S} = 1 = \frac{3g_0}{4} + \frac{g_1}{4}.
\ee

If a strong, saturating LTD protocol is applied to this state, we
reach $\P = (1,0,0)$, where the normalized AMPA conductance is
$g_0$. It has been observed (O'Connor et al.) that after the
induction $\frac{G_{AMPA}}{N_S} = 0.65 \pm 0.03$. We take this to
be $\frac{G_{AMPA}}{N_S} = \frac{2}{3}= g_0$ which implies $g_1 =
2$.

Next apply a phosphatase blocker (okadaic acid was used by O'Connor
et al.) so $g = 0$ and, as they did, present a saturating LTP
signal to arrive at the state $\P =
\frac{(0,af,bf)}{(a+b)f}=\frac{(0,a,b)}{(a+b)}$, which is independent of the
transition rate $f$. This is precisely the fixed point noted
above with $g = 0$ and $h = a f$. After this protocol, the
normalized AMPA conductance is approximately 2, leading to
\be
\frac{ag_1 + bg_2}{a+b} = 2,
\ee
and thus $g_2 = 2$.
Our model corresponds to the set of normalized individual level conductances
 $(g_0 = \frac{2}{3},g_1 = 2,g_2 = 2)$.

The constants $a$ and $b$ are not determined by the observations
so far. Presumably they can be determined by applying
various induction protocols once we have a model for the transition rates.
 The actual time series of
\be
\frac{G_{AMPA}(t)}{N_S} = \frac{2}{3} p_0(t) + 2 p_1(t) + 2p_2(t)
\ee
will depend on $a$ and $b$. O'Connor et al. provide some evidence
that the ratio $\frac{a}{b}$ could be about $\frac{1}{4}$, but it
is not clear saturating protocols were used in these experiments.

This ratio $\frac{a}{b}$ can be determined by another experiment
not yet conducted. Start with the naive synapse $\P(0) =
(\frac{3}{4}, \frac{1}{4},0)$, apply okadaic acid, so $g = 0$, (as in O'Connor
et al.) which blocks phosphatases, and present a saturating LTP
protocol. This leads to the state $\P = \frac{(0,a,b)}{a+b}$.
Now wash out the okadaic acid and apply the kinase blocker k252a,
setting $f = 0$, and present a saturating LTD protocol. This leads
one to the state $\frac{(a,0,b)}{a+b}$. The normalized AMPA
conductance in this state is \be \frac{G_{AMPA}}{N_S} = \frac{ag_0 +
bg_1}{a+b} = \frac{\frac{2}{3}\frac{a}{b}+2}{1 + \frac{a}{b}}. \ee
Measuring $\frac{G_{AMPA}}{N_S}$ after this protocol sequence would give us a
value for $\frac{a}{b}$. If $\frac{a}{b}$ were 0.25, as suggested
above, then this normalized conductance would be $\frac{26}{15}
\approx 1.73$.

\subsection{Transition Rates in a Model for Voltage\\ and Calcium Dynamics}

Evaluation of the transition rates $f$ and $g$ requires a specific model describing
how the postsynaptic cell responds to various induction protocols
presented either presynaptically or as paired presynaptic and
postsynaptic actions. It also requires a model for the dynamics of
\ca in the postsynaptic cell. We proceed using the idea that
changes in AMPA conductance are induced by the time course of
elevation of intracellular \ca ~\cite{Artola,Malen,Sabat,Sjs,Solder}.
 The details of the pathways which follow elevation of
\ca are not specified in the phenomenological approach we use.

In this regard we have explored both one and two compartment models
of the voltage and intracellular calcium dynamics of a cell with
AMPA receptors whose strength is changed as a result of biochemical
pathways activated by the induction protocols. The two compartment
model, which we utilize here, separates the cellular dynamics into
a somatic compartment where action potentials are generated by the
familiar sodium and potassium currents and a dendritic spine
compartment where AMPA and NMDA receptors are located and
intracellular calcium dynamics occurs. The details of this model
are located in the Appendix to this paper. As one improves this
model or replaces it with further experimental insights into the
processes involved in LTP/LTD induction, one can use those
improvements to provide evaluations for the transitions rates
needed in the discrete state synapse model.

The output from the biophysical model of the neuron which we require in this section 
is focused on the time course of elevation
of intracellular calcium concentration relative to the equilibrium concentration 
$C_0 \approx 100\,$nM. We call this time course
$Ca(t) = [Ca^{2+}]_i(t)$ and seek the way in which
\be
\Delta Ca(t) = \frac{Ca(t) - C_0}{C_0}
\ee
influences the transition rates $f(t),g(t)$.

Our model involves two auxiliary variables, $P(t)$ and $D(t)$,
which satisfy first order kinetics driven by Hill functions
dependent on $\Delta Ca(t)$. These variables satisfy
\bea
\frac{dP(t)}{dt} &=& F_P(\Delta Ca(t))(1-P(t)) - \frac{P(t)}{\tau_P} \nonumber \\
\frac{dD(t)}{dt} &=& F_D(\Delta Ca(t))(1-D(t)) - \frac{D(t)}{\tau_D},
\eea
with driving terms
\be
F_P(x) = \frac{x^L}{\xi_P^L + x^L}; \; \; \; F_D(x) = \frac{\alpha x^M}{\xi_D^M + x^M}.
\ee
We used the constants $\tau_P = 10 \,$ ms, $\tau_D = 30\,$ms,
$\alpha = 1.5, L = 8, M = 4.75, \xi_P = 5.5,$ and $\xi_D = 10.5$ in
our calculations for this work. These equations are discussed in
our earlier work ~\cite{Abar1}.

These kinetic quantities are driven by elevation in \ca, $\Delta
Ca(t) > 0$, from their resting value of zero. They are taken to be
related to the transition rates as
\bea
f(t,\Delta Ca(t)) &=& P(t) D(t)^{\eta} \nonumber \\
g(t,\Delta Ca(t)) &=& P(t)^{\eta} D(t),
\eea
and $\eta = 4$ as used in our earlier work. The quantities $f(t)$
and $g(t)$ have dimensions of frequency. Our arguments do not
establish their magnitude but only provide a connection to their
dependence on elevation of intracellular \ca levels. Multiplying
the relations here between $f(t)$ and $g(t)$ and $P(t)$ and $D(t)$
by a constant rescales the time while not affecting the final
states which lead to specific statements of AMPA conductance
changes after an induction protocol.

The model for voltage dynamics and \ca dynamics is now established.
To proceed we specify an electrophysiological protocol. For example
we present a burst of spikes to the presynaptic terminal with an
average interspike interval (ISI) of our choice. Our presynaptic
terminal represents the population of terminals from presynaptic
neurons onto a postsynaptic neuron. This induces a voltage and
\ca response in the postsynaptic cell, and from the time course of
$\Delta Ca(t)$ we evaluate the transition rates $f(t)$ and $g(t)$.
These enter the `master' equation for the average occupations
across the population of $N_S$ synapses. Solving the equations for
the $p_l(t)$ leads to our evaluation of
\be
\frac{G_{AMPA}(t)}{N_S} = g_0p_0(t) + g_1p_1(t)+ g_2p_2(t).
\ee

\subsection{LTP and LTD Induction Protocols}

\subsubsection{Presynaptic Bursts}

The first protocol we used presented a burst of ten spikes to the
presynaptic terminal and evaluated $\frac{G_{AMPA}(t)}{N_S}$ over and
at the end of the induction period. The interspike interval (ISI)
was constant in the burst, and we show in Figure 3 the value of
$\frac{G_{AMPA}(t)}{N_S} -1$ after the burst as a function of frequency
equal to $\frac{1}{ISI}$. Three calculations are presented. The
first, shown with filled circles, involves the action of both the
LTP inducing transition rate $f(t)$ and the LTD inducing
transition rate $g(t)$.  As in the experimental data there is a
region of no change in AMPA conductance per synapse for very low
frequencies, then a region of LTD until this crosses into a region
of persistent LTP. The second calculation, shown with upright
triangles, removes the LTD inducing transition rate, so $g(t) =
0$, which is achieved by O'Connor et al. by the use of okadaic
acid. In this calculation we see that LTP alone is induced at all
frequencies where there is a measurable effect. The maximum AMPA
conductance in the present model is $\frac{G_{AMPA}(t)}{N_S} = 2$
occurring when the lowest state is totally depleted. The value of
unity for $\frac{G_{AMPA}(t)}{N_S} -1$ is expected when a saturating
LTP protocol is applied. Finally, a third result shown in Figure 3
is the set of points with inverted triangles which occur when one
blocks kinase action, again following the experimental procedures
of O'Connor et al., which means $f(t) = 0$ in our language. Here
we see a persistent LTD dropping to $\frac{G_{AMPA}(t)}{N_S} - 1
\approx -\frac{1}{3}$ above frequencies of 10 Hz. This is the
smallest possible value in the present model, as with this induction
protocol and $f(t) = 0$ the lowest state is fully populated, and
the AMPA conductance, in dimensionless, normalized units is
$\frac{2}{3}$.

All of this is consistent with the observations and the expectation
of saturating LTP/LTD protocols in the discrete state plasticity
model we have developed. It is important to note that the bounded
nature of the AMPA conductance is quite important as in many other
models, including our own ~\cite{Kar,Shouval,Abar1}, there is no guarantee that
$\frac{G_{AMPA}(t)}{N_S}$ is bounded above or below.

\subsubsection{Spike Timing Plasticity}

The exploration of spike timing dependent plasticity at hippocampal
synapses has resulted from the investigations of the phenomenon
since the work of ~\cite{Deban,Markram} and
Poo and his colleagues ~\cite{BiPoo1,Nishi,BiPoo2} over the past few years.
 We explored this in the
present model by first presenting a spike presynaptically at a time
$t_{pre}$ and evoking a postsynaptic spike at $t_{post}$. The
change $\frac{G_{AMPA}(t)}{N_S}-1$ is a function only of $\tau =
t_{post} - t_{pre}$ and for our model is shown in Figure 4. This
reproduces the characteristic window of LTP centered near $\tau =
0$ and of width $\approx 10\,$ms around this point. Also shown in
Figure 4 are the LTD regions on both sides of this window. The one
for $\tau$ negative is seen in many experiments. The LTD region for
$\tau$ positive has been seen in experiments reported by ~\cite{Nishi}, and it is common 
in models, including ours, which
focus on postsynaptic intracellular \ca as inducing the chain of
events leading to AMPA plasticity.

~\cite{Nishi} used cesium instead of potassium in the
intracellular pipette solution, and this has been argued to
depolarize the postsynaptic cell and broaden the action potential
artificially ~\cite{Witten}. To address this, ~\cite{Witten}
has performed experiments in which this additional depolarizing
effect is mimicked by presenting a spike timing protocol with one
presynaptic spike at time $t_{pre}$ and \textit{two} postsynaptic
spikes with a time difference $\Delta t$. ~\cite{Witten} uses
$\Delta t = 10$ ms, and the outcome of this protocol for our model
is plotted in Figure 5 with the experimental data (~\cite{Witten};
used with permission). The change in normalized synaptic strength
resulting from this protocol ($G_{AMPA}/N_{S}$ - 1 for the model)
is shown as a function of the time of the second postsynaptic spike
$t_{post (2)} - t_{pre}$. For the experiments, normalized synaptic
strength is the average peak excitatory postsynaptic current (EPSC)
height measured 10-20 minutes after the end of the pairing
protocol, normalized by the mean baseline peak EPSC height. It is
clear that the LTP window is substantially larger than when we
evoke just one postsynaptic spike and resembles the experimental
data. We can regard this as a prediction of our discrete state
plasticity model. Further predictions of this protocol are shown in
Figures 6 and 7 where $\Delta t = 15\,$ms and $\Delta t = 20\,$ ms
respectively. In each case there is a distinct LTD window for
positive $t_{post (2)} - t_{pre}$ and a distinctive dip between the
LTP peaks whose separation is dictated by $\Delta t$.

\subsection{Synchronization of Two Periodic Neural Oscillators with Discrete State Synapses}

The final consequence we have investigated of our discrete state
plasticity model is for the synchronization of oscillating neurons.
We take as given that synchronization among populations of neurons
can play an important role in their performing important functional
activity in biological neural networks. We have abstracted the
synchronous activity of populations of neurons to the simplest
setup: two periodically oscillating Hodgkin-Huxley (HH) neurons
coupled by a synaptic current which we explore with and without
plastic synapses.

We have selected the postsynaptic neuron to be our two compartment
model as described in the Appendix and set it into autonomous
oscillations with a period $T_2^0$. This period is a function of
the injected DC current into the somatic compartment. We hold this
fixed while we inject a synaptic AMPA current
\be
I_{\mbox{synapse}}(t) = g_{AMPA}(t) S_A(t) (E_{rev} - V_{post}(t)),
\ee
into the postsynaptic somatic compartment. $V_{post}(t)$ is the
membrane voltage of this postsynaptic compartment. $g_{AMPA}(t)$ is our
time dependent maximal AMPA conductance, and $S_A(t)$ satisfies
\be
\frac{dS_{A}(t)}{dt} = \frac{1}{\tau_A} \frac{S_{0}(V_{pre}(t))-S_{A}(t)}{S_{1A}- S_{0}(V_{pre}(t))}
\ee
as described in detail in the Appendix. $V_{pre}(t)$ is the
periodic presynaptic voltage which we adjust by selecting the
injected DC current into the presynaptic HH neuron. We call the
period of this oscillation $T_1$.

When $g_{AMPA} = 0$ the neurons are disconnected and oscillate
autonomously. When $g_{AMPA}(t) \ne 0$ the synaptic current into
the postsynaptic neuron changes its period of oscillation from the
autonomous $T_2^0$ to $T_2$, which we evaluate for various choices
of $T_1$. We expect from general arguments ~\cite{Dranzin} that there
will be regimes of synchronization where $\frac{T_1}{T_2}$ equal
integers and half-integers over the range of frequencies
$\frac{1}{T_1}$ presented presynaptically. This will be true both
for fixed $g_{AMPA}$ and when $g_{AMPA}$ varies as dictated by our
model.

In Figure 8 we present $\frac{T_1}{T_2}$ as function of the
frequency $\frac{1000}{T_1}$ ($T_1$ is given in milliseconds, so
this is in units of Hz) for fixed $g_{AMPA} = 0.1 \frac{mS}{cm^2}$ and
for $g_{AMPA}(t) = g_{AMPA} \frac{G_{AMPA}(t)}{N_S}$ determined from our model.
This value is what we used in our earlier calculations with the two
compartment model. It amounts to a choice for the baseline value of
the AMPA conductance.
The fixed $g_{AMPA}$ results are
in filled upright triangles and, as expected, show a regime of
one-to-one synchronization over a range of frequencies. One also
sees regions of two-to-one and hints of five-to-two and
three-to-one synchronization. These are expected from general
arguments on the parametric driving of a nonlinear oscillator by
periodic forces.

When we allow $g_{AMPA}$ to change in time according to the model
we have discussed above, we see (unfilled inverted triangles) a
substantial increase in the regime of one-to-one synchronization,
the appearance of some instances of three-to-two synchronization,
and a much smaller regime with two-to-one synchronization. This
suggests that the one-to-one synchronization of oscillating
neurons, which is what one usually means by neural synchrony, is
substantially enhanced when the synaptic coupling between neurons
is allowed to vary by the rules we have described.

We show the same results in Figure 9 for $g_{AMPA} = 0.2 \frac{mS}{cm^2}$.
The fixed
coupling is larger leading to stronger synchronization of the two
neurons in a one-to-one  manner even for fixed coupling. Here too
(inverted, unfilled triangles) we see that allowing $g_{AMPA}$ to vary
in time enlarges the regime of one-to-one synchronization.

In Figures 10 and 11 we explore aspects of the internal dynamics of
plasticity and $\ca$ time courses for these results. In Figure 10
we show $Ca(t) = [Ca^{2+}]_i(t)$ (scaled by a factor of 15 to fit
on this graphic) and $\frac{G_{AMPA}(t)}{N_S} -1$ in response to a
presentation of periodic presynaptic oscillations beginning at a
time 300. As noted earlier, the timescales for the intracellular
\ca processes and the timing in changes in $G_{AMPA}(t)$ are not
determined by our model. An arbitrary constant can multiply the
definitions of the transitions rates $f(t)$ and $g(t)$. Both
quantities rapidly rise, after a small transient of LTD, to
positive but oscillating levels. The maximum
$\frac{G_{AMPA}(t)}{N_S}
-1$ is 1 in our model, and we see that this saturating level is not
reached in this protocol.

Finally, in Figure 11 we examine how the synchronization manifests
itself in the postsynaptic somatic and dendritic compartment
membrane potentials. We plot these potentials along with
$V_{pre}(t)$. It is clear that the one-to-one synchronization
occurs with an in-phase oscillation of the presynaptic and
postsynaptic cells. The very short time delay between the somatic
and dendritic compartments of the postsynaptic neuron is part of
the model dynamics and not associated with the presentation of
periodic presynaptic spikes to the postsynaptic cell. The in-phase
synchronization is not seen in other, less biophysically based,
models of plastic synapses and represents a very desirable feature
of this model.

\section{Discussion}

The observations, recent and over the years, of discrete levels for
synaptic strength at individual synapses in the CA3-CA1 hippocampal
pathways represents a fundamental property important for the ways
we learn and remember. There is a very interesting and important
biophysical question about the mechanisms which lead to the
expression of a few discrete levels of AMPA conductance at which
individual synapses may be found. We do not address this
fundamental question in this paper, but we have used the
observation to construct a model based on discrete levels with
transition rates among the levels determined by biophysical
dynamics.

We have formulated a discrete level synaptic system in a general
way with $L$ levels allowed to the AMPA conductance, and then,
following the observations of O'Connor et al., we  specialized to
$L = 3$. The dynamical variables in our model when $L = 3$ are the average
occupation numbers of each level $\P(t)=[p_0(t),p_1(t),p_2(t)]$.
These are averages over a collection of $N_S$ synapses which
contribute to the overall AMPA determined response of the neuron.
While each individual synapse resides in one of three discrete
states, so the individual occupation numbers at any given synapse
are either zero or one, the average occupation numbers are smoothly
varying, subject only to $p_0(t) + p_1(t) + p_2(t) = 1$, by
definition.

We developed differential equations for $\P(t)$ which are linear in
the $p_l(t),\; l = 0,1,2$
\be
\frac{dp_l(t)}{dt} = \sum_{l'=0}^2 M_{ll'}p_{l'}(t),
\ee
and where the transition rates $M_{ll'}$ are determined by nonlinear membrane voltage and 
intracellular \ca dynamics.

From the observations of O'Connor et al. we argued that the
transition rates shown in Figure 1 sufficed to explain their
measurements, and using their reported results we were able to
determine that the conductances of the three individual levels in
normalized, dimensionless units were $g_0 = \frac{2}{3}, g_1 = 2,
g_2 = 2$. The time dependence of the normalized, dimensionless AMPA
conductance per synapse is then
\be
\frac{G_{AMPA}(t)}{N_S} = \sum_{l=0}^2\,p_l(t)g_l.
\ee

Using our values for the $g_l$ and a dynamical model of the
transition rates $f(t),g(t)$ as shown in Figure 1, we reproduced
the observed plasticity in response to a burst of presynaptic
spikes with interspike intervals (ISIs) over the observed range.
Further we made predictions for the response of this model to spike
timing plasticity both for one presynaptic and one postsynaptic
spike and for the case of two postsynaptic spikes evoked $\Delta t$
apart accompanied by one presynaptic spike. We presented our
results for $\Delta t = 10, 15, ..., 20\,$ms.

Finally we examined the dynamical role played by this discrete
state plasticity model in the synchronization of two periodically
oscillating Hodgkin-Huxley neurons. One such neuron oscillating
with period $T_2^0$ was driven by another such neuron with period
$T_1$. The final period $T_2$ of the driven neuron, relative to
$T_1$, was plotted against $\frac{1}{T_1}$ and showed familiar
regions of synchronization. For fixed AMPA coupling $g_{AMPA} =
\frac{G_{AMPA}}{N_S}$ we found synchronization over some range of
$\frac{1}{T_1}$ and then demonstrated that allowing $g_{AMPA}$ to
vary according to the plasticity model resulted in a much larger
regime of one-to-one synchronization with the two neurons
oscillating in-phase. The results for synchronization have not been
tested experimentally, though some experiments using dynamic clamp
based synapses have been performed.

One striking aspect of the discrete state model, certainly not
limited to our own work, is that the AMPA conductance has natural
upper and lower bounds. Many other models of plasticity, including
our own, do not share this important feature.

Some of our results, in particular the strengths of the normalized,
dimensionless conductances of the synaptic levels are dependent
primarily on the data of O'Connor et al. All of the transition
rates are determined by our two compartment model for the neuron,
as presented in the text and in the Appendix.

Our model exhibits a number of features of LTP and LTD observed
experimentally at CA3-CA1 synapses, including trapping of synapses
in a high-strength state, separability of potentiation and
depression by simulated inhibition of kinase or phosphatase
activity, and spike timing-dependent plasticity (O'Connor et al.;
~\cite{Witten}). However, there are a number of ways in which the
model can be developed further. For example, using the protocols of
O'Connor et al., population LTP at CA3-CA1 synapses rises gradually
to a peak level over a few minutes; LTD takes a few minutes longer
than this to develop fully. Our model does not yet include such a
long timescale, but could be modified phenomenologically to do so.
O'Connor et al. have also found that the high locked-in state in
populations of synapses builds up over several minutes. To model
this phenomenon, we would again need to include a longer timescale.

While the spike timing-dependent plasticity induced in our model by
1 presynaptic and 2 postsynaptic spikes is similar to that observed
by ~\cite{Witten}, our result for 1 presynaptic and 1
postsynaptic spike appears to differ from experimental observations
(G. M. Wittenberg and S. S.-H. Wang, unpublished data used with the
authors' permission). In particular, they observed little LTP but
significant LTD near $t_{post}$ - $t_{pre}$ = 0. At other values of
$t_{post}$ - $t_{pre}$, they observed only LTD. This suggests that
such a protocol provides insufficient postsynaptic \ca influx to
induce LTP reliably. In contrast, our model shows a narrow but
clear window of LTP centered near $t_{post}$ - $t_{pre}$
= 0. At these values of $t_{post}$ - $t_{pre}$, the \ca influx in
our model is sufficient to give f a relatively large value and thus
induce LTP. If the data of Wittenberg and Wang are correct, then
our model will need to be adjusted so that this \ca influx is not
sufficient to induce LTP.

In addition, the model can be refined to match the results of
various LTP and LTD induction protocols that we have not simulated
here but that are used by O'Connor et al. and others in their
experiments, like theta burst stimulation and pairing protocols. In
the long term, a model that more accurately describes the
postsynaptic signaling pathways will eventually account for all of
these various features of the data in a biologically satisfying
manner.

While our model is based on the idea of discrete state synapses, by
design it describes only populations of such synapses. In future
work we plan to address this discreteness at the level of single
synapses and small numbers of synapses. In a single-synapse model,
transition probability functions would likely replace the
transition rate functions we have described in the present model,
but many other features of the model could be retained.

This model will change over time and be improved by further
understanding of the biophysical processes leading to the discrete
states and their transitions among themselves. The general
framework we have presented describing how the three observed
states are connected and several general results about that system
will remain as the representation of the transition rates is
improved.
\bigskip
\bigskip

{\bf Acknowledgments} This work was partially funded
by the U.S. Department of Energy, Office of
Basic Energy Sciences, Division of Engineering and Geosciences,
under Grants No. DE-FG03-90ER14138 and No. DE-FG03-96ER14592; by a
grant from the National Science Foundation, NSF PHY0097134,
and by a grant from the National Institutes of Health, NIH R01 NS40110-01A2.
We have had many productive conversations with S. S.-H. Wang about the
data from his laboratory, and we appreciate the comments he has made
on the model presented here as well as the ability to see his results
before publication.

\clearpage
\section{Appendix}

In this Appendix we give the details of the two compartment model
of the postsynaptic neuron used in our calculations. The somatic
compartment is the site of spike generation by the familiar
Hodgkin-Huxley (HH) \na, \k, and leak currents. The dendritic
spine compartment has these currents as well as a voltage gated
calcium current and glutamate driven NMDA and AMPA channels. The
\ca dynamics in the dendritic compartment drives the synaptic
plasticity, namely changes in the AMPA conductance.

\subsection{The Somatic Compartment}

 The dynamical equation for the somatic compartment takes the general form,
 \bea
 C_M \frac{dV_{S}(t)}{dt} &=& I_{Na}(V_S(t),t) + I_{K}(V_S(t),t) + I_{L}(t) \nonumber \\
 &+& I_{Sdc} +I_{S}(t) \nonumber\\
 &+& G_{S \leftarrow D} (V_{D}(t)-V_{S}(t))
 \eea
The currents $I_{Na}$, $I_{K}$, $I_{L}$, are the familiar HH \na,
\k, and leak currents. $I_{Sdc}$ is a DC current used to set the
resting potential of the cell. $I_{S}(t)$ is an externally managed
time dependent current injected into the somatic compartment. It
allows us to evoke an action potential at a specific time in the
somatic compartment. This propagates back to the dendritic
compartment to induce a depolarizing effect. $G_{S \leftarrow D}
(V_{D}(t)-V_{S}(t))$ represents the current flowing into the
somatic compartment from the dendritic compartment. It couples the
voltages of the somatic and dendritic compartments. $C_M$ is the
membrane capacitance.

The value of the currents $I_{Na}$, $I_{K}$, and $I_{L}$ are
determined as usual with
\be
I_L(t) = g_L(E_L - V_S(t)),
\ee
where $g_L$ is the conductance of the leak current and $E_L$ is the
reversal potential. The voltage gated currents are described by
\be
I(V,t) = \bar{g}g(t,V)(E_{eq}-V),
\label{ohm}
\ee
 where $E_{eq}$ is the reversal potential and $\bar{g}$ is the
 maximal conductance. Both these values are fixed. The value of
 $g(t,V)$, the fraction of open channels, on the other hand depends on the membrane potential
 and time.

In the case of channels in which $g(t,V)$ changes, the value of
$g(t,V)$ depends on the state of `gating particles' $m(t,V)$ and
$h(t,V)$, where $m(t,V)$ is the activation gate and $h(t,V)$
represents the  inactivation gate. If $N$  is the number of
activation gates, and $M$, the number of inactivation gates then

$$g(t,V)=m(t,V)^Nh(t,V)^M$$

The state of the gating particles, given by $m(t,V)$ or $h(t,V)$,
is a function of the membrane potential as well as time.

These gating variables, denoted by $X(t)$, are taken to satisfy
first order kinetics.
 \begin{eqnarray}
 \frac{dX(t)}{dt} &=& \frac{X_{0}(V(t)) - X(t)}{\tau_{X}(V(t))} \nonumber \\
                  &=& \alpha_{X}(V(t))(1-X(t))-\beta_{X}(V(t))X(t)
 \end{eqnarray}
From the standard HH model, we have the following relations for the
conductances of the $Na^{+}$ and $K^{+}$ currents.
\bea
 g_{Na}(V,t) &=& m(V,t)^3h(V,t) \nonumber \\
 g_{K}(V,t) &=& n(V,t)^4 \nonumber
\eea
where $m(V,t),n(V,t)$ are activation gating particles and $h(V,t)$
represents the inactivation gating particle. At the end of this
appendix we give the functions $X_{0}, \tau_{X},$ or
$\alpha_{X}(V),
\beta_{X}(V)$ for each of the voltage gated ionic currents, in
addition to listing all the model parameters used in the simulations.
\subsection{The Dendritic Compartment}
The dynamics of the dendritic compartment membrane potential is
given by
\bea
C_{M}\frac{dV_{D}(t)}{dt} &=& I_{Na}(V_D(t),t) + I_{K}(V_D(t),t) + I_{L}(V_D(t)) \nonumber \\ 
&+& I_{A}(V_D(t),t)+ I_{M}(V_D(t),t) + I_{Ddc}\nonumber \\
&+& I_{AMPA}(t) +I_{NMDA}(t) + I_{VGCC}(t) \nonumber \\ &+& G_{D
\leftarrow S} (V_{S}(t) - V_{D}(t)) \eea $I_{Ddc}$ is a DC current
used to set the resting potential of the cell. $I_{Na}, I_{K}$ and
$I_{L}$ represent the standard HH ionic and leak currents used in
the somatic compartment as described above. In addition to these
ionic currents we have considered two additional $K^{+}$ currents,
$I_{A}$ and $I_M$. $I_{A}$ currents have been reported to modulate
the width of action potentials and influence the excitability of
the cell. In our model, $I_{A}$ attenuates the dendritic action
potential, which is evoked by backpropagation of the somatic action
potential.

The gating equations for $I_{M} $ and $I_{A}$ are,
$$g_{M}(t,V)=u(t,V)^2$$
 $$g_{A}(t,V) = a(t,V)b(t,V)$$
where $u(t,V)$ and $a(t,V)$ are activation gating particles and
$b(t,V)$ represents an inactivation gating particle.

As mentioned above all ionic currents in the dendritic compartment
are also given in terms of Ohm's law, (\ref{ohm}). Again $G_{D
\leftarrow S} (V_{S}(t) - V_{D}(t))$  represents current flowing
into the dendritic compartment through the somatic compartment.

In addition to the currents mentioned above we have three other
currents critical to the synaptic plasticity discussed here. There
is a current associated with the ligand gated NMDA receptors
(NMDARs). The form for this is
\bea
I_{NMDA}(t)=g_{NMDA}S_{N}(t)B(V_{D}(t))(V_{NMDA-eq} -V_{D}(t))
\eea
where $g_{NMDA}$ is the maximal conductance associated with the channel.
$S_{N}(t)$ ranges between zero and unity, representing the
percentage of open channels at any time. To achieve the time course
of this process in NMDARs, we use a two component form for $S_{N}$,
\be
S_{N}(t)=w_fS_{N1}(t)+(1-w_f)S_{N2}(t) ,
\ee
$0 \le w_f \le 1$, and where $S_{Nl}(t), l=1,2$ satisfies
\be
\frac{dS_{Nl}(t)}{dt} = \frac{1}{\tau_{Nl}}
\frac{S_{0}(V_{pre}(t))-S_{Nl}(t)}{S_{1Nl}- S_{0}(V_{pre}(t))}
\ee
$V_{pre}$ is scaled to lie between 0 and 1 as it represents the
arrival of an action potential at the presynaptic terminal. Its
function is in turn to release neurotransmitter.

$S_{0}(V_{pre}(t))$ is a step function which rises sharply from 0
to 1 when neurotransmitter is released as a result of the
presynaptic action potential. When this occurs $S_{Nl}(t)$ rises
from zero towards unity with a time constant
$\tau_{Nl}(S_{1Nl}-1)$. When the effect of presynaptic action is
completed, $S_{Nl}(t)$ relaxes towards zero with a time constant
$\tau_{Nl}S_{1Nl}$. $w_f$ represents the fraction of fast NMDA
component contribution to NMDA current. In our model we have chosen
$w_{f}=.81$, $\tau_{N1}=67.5$, $S_{1N1}=70/67.5$, $\tau_{N2}=245$,
$S_{1N2}=250/245$. In addition the conductance of the NMDA current
depends on postsynaptic voltage via the term $B(V)$ whose form is
given as,

\bea
B(V) = \frac{1}{1 + .288[Mg^{2+}]e^{-.062V}},
\eea
where the concentration of magnesium is in mM and the voltage is in
mV. For simulation purposes we have taken the physiologically
reasonable value of $[Mg^{2+}] =1\,$ mM.

This voltage dependent conductance depends on the extracellular
magnesium concentration. The voltage dependence of the current is
mediated by the magnesium ion which, under normal conditions,
blocks the channel. The cell must therefore be sufficiently
depolarized to remove the magnesium block. Finally for this
excitatory channel $V_{NMDA-eq} \approx 0\,$mV.

$I_{AMPA}$ represents the ligand gated AMPA receptor current. This
is taken to be of the form,
\be
I_{AMPA} = g_{AMPA}S_{A}(t)(V_{AMPA-eq} - V_{D}(t))
\ee
where $g_{AMPA}$ is the maximal conductance for this channel and
$S_{A}(t)$ is the percentage of open channels, satisfying
\be
\frac{dS_{A}(t)}{dt} = \frac{1}{\tau_A} \frac{S_{0}(V_{pre}(t))-S_{A}(t)}{S_{1A}- S_{0}(V_{pre}(t))}
\ee
Again the rise time is less than a millisecond. In our formulation
this time is $\tau_{A}(S_{1A}-1)$, which we set to 0.1 ms. AMPA
currents decay in approximately 1-3 ms. In our formulation this
decay time is $\tau_{A}S_{1A}$, which we set to 1.5 ms. We also
take $V_{AMPA-eq}
= 0\,$ mV.

The final and very important ingredient in inducing synaptic
plasticity is the voltage gated calcium channel (VGCC). We have
used the low threshold current $I_T$ for this. The current from
this channel takes the form,
\be
I_{VGCC}(t) = g_{C} GHK(V(t))m_{c}^{2}(t)h_{c}(t)
\ee
where $g_C$ is the maximal conductance of this channel, $m_c(t)$ is
the activation function, and $h_{c}(t)$ is the inactivation
function. GHK(V) is the Goldman-Hodgkin-Katz function,
\bea
GHK(V) &=& -\frac{V}{C_0} \frac{[Ca^{2+}]_{i}(t) -
[Ca^{2+}]_{o}e^{\frac{-2VF}{RT}}}{1- e^{\frac{-2VF}{RT}}} \nonumber \\
       &=& - \frac{V}{C_0} \frac{Ca(t) - [Ca^{2+}]_{o}e^{\frac{-2VF}{RT}}}{1- e^{\frac{-2VF}{RT}}}
\eea which is used because of the disparity in the intracellular
$[Ca^{2+}]_i$ and the extracellular $[Ca^{2+}]_o$ concentrations.
$F$ is Faraday's constant, R is the gas constant, and T the
absolute temperature. Other factors of the GHK equation are
absorbed in the conductance $g_{C}$. $C_o$ is the equilibrium
intracellular $[Ca^{2+}]$ concentration, which is about 100 nM.
\subsection{Coupling between the somatic and dendritic compartments}
The coupling parameters between the two compartments are determined
from the cytoplasmic resistance and the metric dimensions of each
compartment. We take the specific cytoplasmic resistance of the
cell to be $r_i =200 \Omega \,$ cm.

We take the somatic compartment to be a isopotential sphere of
$d_{Soma} = 32.5 \mu m$ in diameter and the dendritic compartment
to be an isopotential cylinder of diameter $d_{Dendrite} = 10 \mu
m$ and length $l_{Dendrite} = 360 \mu m$.

In order to determine the coupling resistance value we assume the
somatic compartment to be a cylinder of equivalent surface area. We
then have the total cytoplasmic resistance of the somatic
compartment,
$$
R_{ISoma} =  \frac{r_{i} l_{Soma}}{\mbox{Crosssection} (A_{Soma})} =
.007839 \times 10^7 \Omega
$$
while the total cytoplasmic resistance for the dendritic
compartment is
$$
R_{IDendrite} = \frac{r_{i} l_{Dendrite}}{\mbox{Crosssection}(A_{Den})} =
1.713 \times 10^7 \Omega
$$
Therefore, the average cytoplasmic coupling
resistance $$R_{I} = \frac{R_{ISoma}+ R_{IDendrite}}{2} \approx
\frac{R_{IDendrite}}{2} = .861 \times 10^7  \Omega
$$
The coupling parameters in units of $mS/cm^2$ are calculated as
\bea
G_{S \leftarrow D} = \frac{1}{A_{Soma} R_{I}} = 3.5 mS/cm^2
\eea and
\bea
G_{D \leftarrow S} = \frac{1}{A_{Den} R_{I}} = 1.0 mS/cm^2
\eea
\subsection{Calcium Dynamics}
The dynamics of intracellular calcium $[Ca^{2+}(t)]$ in the dendritic
 compartment, which affects
the efficacy of synaptic strength, is comprised of
 $[Ca^{2+}(t)]$ decaying to an equilibrium value of $C_o$ on
 a timescale of $\tau_{C} \approx 15ms $, which we take to be
 about 30 ms in our model,
  plus fluxes of $[Ca^{2+}(t)]$ due
to the three channels, AMPA, NMDA, and VGCC considered in the
dendrite model above. The first order differential equation for
$[Ca^{2+}(t)]=Ca(t)$  then is
 \begin{eqnarray}
 \frac{dCa(t)}{dt} &=& \frac{1}{\tau_C} (C_{o} - Ca(t)) + Ca_{NMDA} + Ca_{AMPA}\nonumber \\
 &+& Ca_{VGCC}
 \end{eqnarray}
 where
$$Ca_{NMDA} = g_{NC}S_{N}(t)B(V_{den}(t))(V_{NMDA-eq} - V_{den}(t))$$
$$Ca_{AMPA} = g_{AC}S_{A}(t)(V_{NMDA-eq} - V_{den}(t))$$
$$Ca_{VGCC} = g_{CC} GHK(V(t))m_{c}^{2}(t)h_{c}(t)$$
 The constants $g_{NC},g_{AC},g_{CC}$ are not the same, even dimensionally, as the conductances in the
 voltage equation. Their values, given at the end of this appendix, reflect among other things, that the net AMPA current is
 composed primarily of other ions besides $Ca^{2+}$ and that NMDA channels are highly permeable to $Ca^{2+}$ ions.
This completes the description of our model.
\subsection{Parameters in the Two Compartment Model}
The various constants appearing in our two compartment model are collected here.
\begin{itemize}
\item {\bf  The somatic compartment :}
The maximal conductances, in units of $mS/cm^{2}$, of the ionic
currents are, $g_{Na} = 215, g_{K}=43, g_{L}=.813$. The reversal
potentials in units of mV are $V_{Na-eq} = 50, V_{K-eq} = -95,$ and
$V_{L-eq}=-64$. The DC current injected into the somatic
compartment $I_{Sdc}= -7.0 \mu A/cm^{2}$, so that the cell is at
-75mV at rest. The magnitude of the additional current injected into the somatic compartment is $I_{Soma}
= 160.8 \mu A/cm^{2}$; in using it to induce a postsynaptic spike,
it is taken to have a duration of 1 ms or less.
\item {\bf  The dendritic compartment :}
For the standard HH ionic currents we have the same parameters as
above. Maximal conductances associated with various dendritic
currents, in units of $mS/cm^{2}$, are $g_{NMDA} = 0.05,
g_{AMPA}=1.75$ and $g_{C} = 1.
\times 10^{-6}$. In the $Mg^{2+}$ blockage function $B(V)$, we take
$[Mg^{2+}] = 1mM$. In the GHK function, the ratio of external
$[Ca^{2+}]$ to equilibrium intracellular $[Ca^{2+}]$ is 15000. The
temperature is $25^{o}C$. The conductance values for additional
potassium currents used are $g_M =6.7$ and $g_A = 100$ in units of
$mS/cm^{2}$. Finally, $I_{Ddc}= -7.0 \mu A/cm^{2}$
\item {\bf  Calcium dynamics :}
For calcium dynamics we have $\tau_{C} = 30 ms, g_{NC} = .15,
g_{AC} = 1.5 \times 10^{-5}, $ and $g_{CC}= 3.5 \times 10^{-5}$.
These are in units of $mV^{-1} ms^{-1}$. $C_{o}$, the basal calcium
concentration in the cell, is normalized to 1.
\end{itemize}
Finally in the equation for the NMDA and AMPA channel open
percentages $S_{N}(t) $ and $S_{A}(t)$, we use the `step' function
$$S_{0}(V) = \frac{1}{2} (1+ \tanh(120(V-.1)))$$
and the constants $\tau_{A} =
1.4ms$, $\tau_{Nfast} = 67.5 ms$, $\tau_{Nslow} = 245.0 ms$,
$S_{1A} = \frac{15}{14}$ , $S_{1Nfast} = \frac{70}{67.5}$,
$S_{1Nslow} =\frac{250}{245} $. 
\subsection{Activation and deactivation parameters of various channels}

\begin{tabular}{c r}

$\alpha_{m}(V) = \frac{.32(13-(V-Vth))}{e^{ \frac{(13-(V-Vth))}{4.0}}-1}$ 
 &
 $\beta_{m}(V) =  \frac{0.28((V-Vth)-40)}{e^{\frac{((V-Vth)-40)}{5}}-1}$ \\ \\
$\alpha_{h}(V) = .128 e^{\frac{17-(V-Vth)}{18}} $ &
$\beta_{h}(V) = \frac{4}{e^{\frac{40-(V-Vth)}{5}}+1}$ \\ \\
$ \alpha_{n}(V) = \frac{0.032(15-(V-Vth))}{e^{\frac{(15-(V-Vth))}{5}}-1}$ &
$ \beta_{n}(V) = \frac{0.5}{e^{\frac{(V-Vth)-10}{40}}}$ \\ \\
$m_{co}(V) = \frac{1}{1+ e^{-\frac{(52+V)}{6.2}}} $ &
$\tau_{mc}(V) = .204 + \frac{0.333}{e^{-\frac{(131+V)}{16.7}} + e^{\frac{(15+V)}{18.2}}} $\\ \\
$h_{co}(V) = \frac{1}{1 + e^{\frac{(72+V)}{4}}} $ &
$\alpha_{u}(V) = \frac{0.016}{e^{-\frac{(V+52.7)}{23}}} $ \\ \\
$\beta_{u}(V) = \frac{0.016}{e^{\frac{(V+52.7)}{18.8}}}  $ &
$ \alpha_{a}(V) = \frac{-0.05(V+20)}{e^{-\frac{V+20}{15}}-1} $\\ \\
$ \beta_{a}(V) = \frac{0.1(V+10}{e^{\frac{(V+10)}{8}}-1} $ &
$ \alpha_{b}(V) = \frac{0.00015}{e^{\frac{(V+18)}{15}}}$ \\ \\
  $\beta_{b}(V) = \frac{0.06}{e^{-\frac{(V+73)}{12}}+1}$
          \end{tabular}
\bea
\tau_{hc}(V) = 0.333e^{\frac{(V+466)}{66.6}} \;\mbox{if} \, V \leq -81  \nonumber \\
          = 9.32 + 0.333e^{-\frac{(V+21)}{10.5}} \;\mbox{if}\, V > -81  \nonumber 
\eea
where $Vth=-48\, $mV.\\

\clearpage

\section{Figures}
\begin{figure}[ht!]
\includegraphics[width=3in,scale=.35]{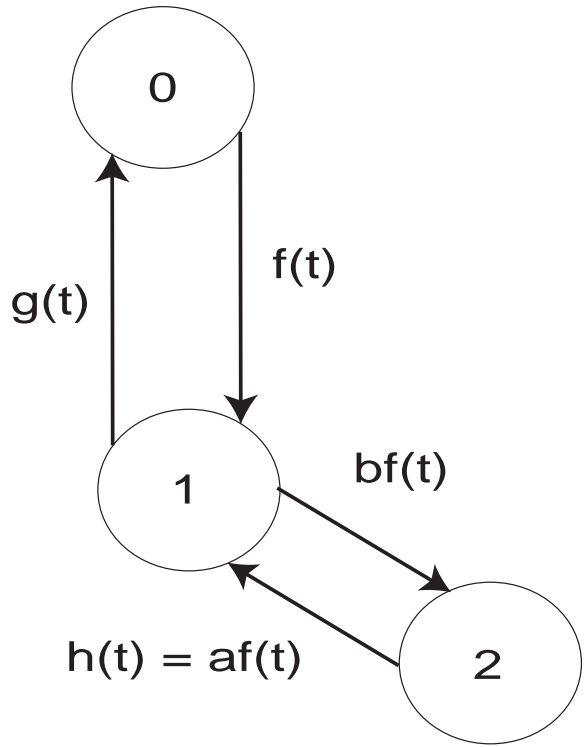}
\caption{ Three state model of synaptic plasticity. Individual
synapses can move among the three states, marked 0 for the ``low" state,
1 for the ``high" state, and 2 for the ``high locked-in" state. The rules for
state transition depend on the transition rates, $f(t)$ and $g(t)$,
governed by changes in intracellular calcium concentration. These
transition rates are determined by LTP and LTD induction protocols.
\label{fig 1}}
\end{figure}

\begin{figure}[ht!]
\includegraphics[width=3in,scale=.35, angle=-90]{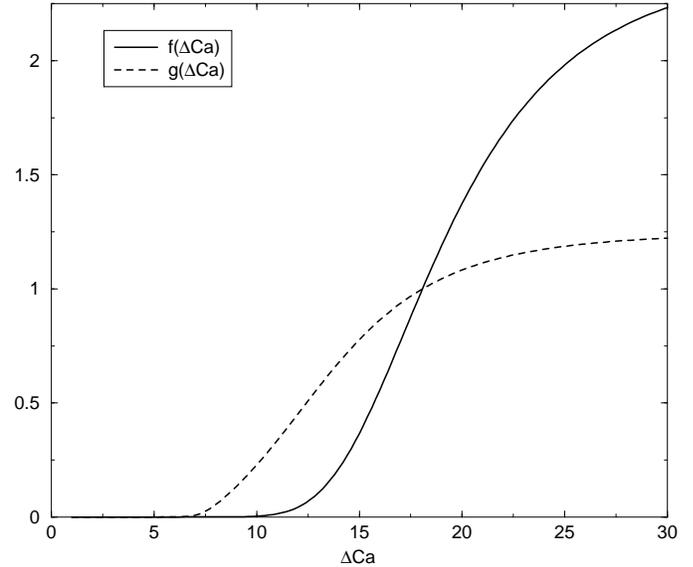}
\caption{ Steady state values for the transition rates $f$ and
$g$ plotted as function of the magnitude of the change in intracellular
calcium concentration. $\Delta Ca$ is in arbitrary units. For small
changes in intracellular calcium concentration only $g$ is
nonzero, corresponding to LTD induction, and for larger changes
in intracellular calcium concentration, both $f$ and $g$ are
nonzero with $f$ being greater than $g$,
corresponding to an LTP induction protocol.
\label{fig 2}}
\end{figure}

\begin{figure}[ht!]
\includegraphics[width=2.8in,scale=.35,angle=-90]{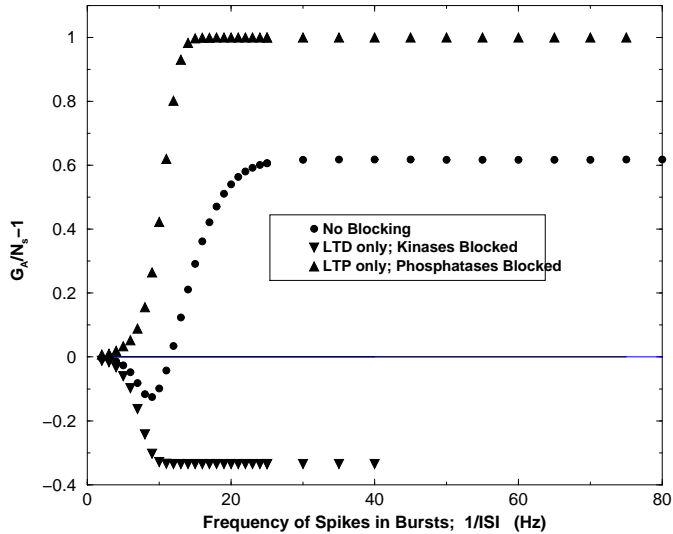}
\caption{Frequency-plasticity curve. The change in
normalized AMPA conductance per synapse, $G_{AMPA}/N_{s}-1$, is
plotted as function of the frequency of a periodic burst of 10
presynaptic spikes presented to the presynaptic terminal. The
circles represent synaptic plasticity for the full three state
model. The upward-pointing triangles represent synaptic plasticity
with the term $g(t)$ set to 0, corresponding to blocking
phosphatase activity in the postsynaptic cell. One sees and expects
LTP alone. The downward-pointing triangles represent the change in
synaptic plasticity with the term $f(t)$ set to 0, corresponding to
blocking kinase activity in the postsynaptic cell. We observe and
expect LTD alone in this case. These results are quite similar to
the observations of O'Connor et al. 
\label{fig 3}}
\end{figure}

\begin{figure}[ht!]
\includegraphics[width=2.8in,scale=.35,angle=0]{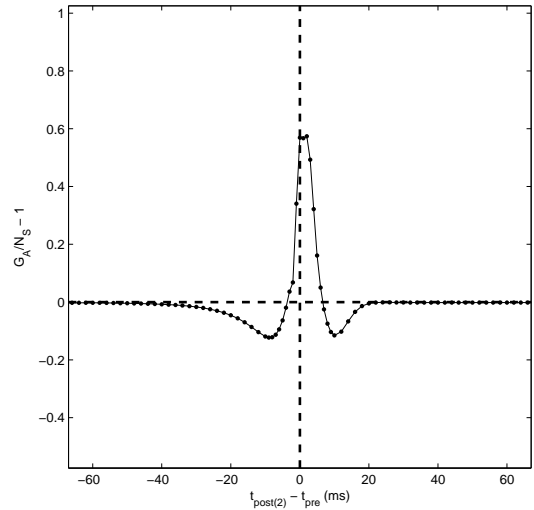}
\caption{ Spike timing dependent plasticity protocol. The change in
normalized AMPA conductance per synapse, $G_{AMPA}/N_{S}-1$, plotted as a function
of the delay, $\tau=t_{post}-t_{pre}$ (ms), between presentation of a single
presynaptic spike at $t_{pre}$ and postsynaptic spike at $t_{post}$.
\label{fig 4}}
\end{figure}

\begin{figure}[ht!]
\includegraphics[width=3in,scale=.35]{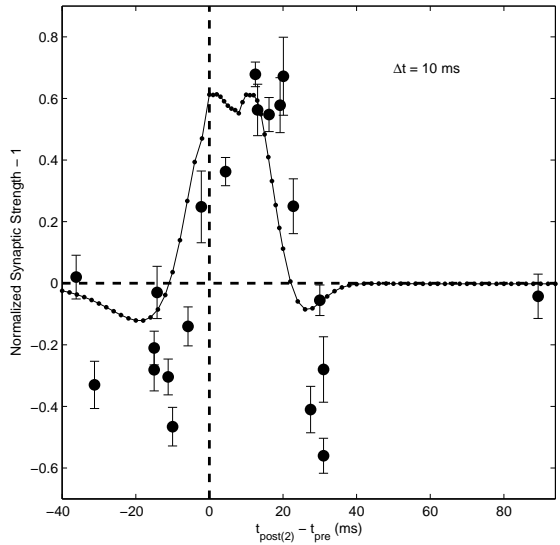}
\caption{ Change in normalized synaptic strength as a
function of the delay $\tau=t_{post(2)}-t_{pre}$ when a single
presynaptic spike is paired with two postsynaptic spikes 10 ms
apart. $t_{post(2)}$ is the time of the second postsynaptic spike.
Model results (\textit{points connected by lines}) are plotted with
experimental data of G. M. Wittenberg (\textit{large filled circles
with error bars}; Wittenberg 2003, used with permission).
\textit{Normalized synaptic strength}, for the model, is the
normalized AMPA conductance per synapse ($G_{AMPA}/N_{S}$) after
the pairing. For the experiments, it is the average peak excitatory
postsynaptic current (EPSC) height measured 10-20 minutes after the
end of the pairing protocol, normalized by the mean baseline peak
EPSC height (\textit{Error bars}: standard error of the mean). In
the experiments, pairing was repeated 100 times at 5 Hz.
\label{fig 5}}
\end{figure}

\begin{figure}[ht!]
\includegraphics[width=3.in,scale=.35,angle=0]{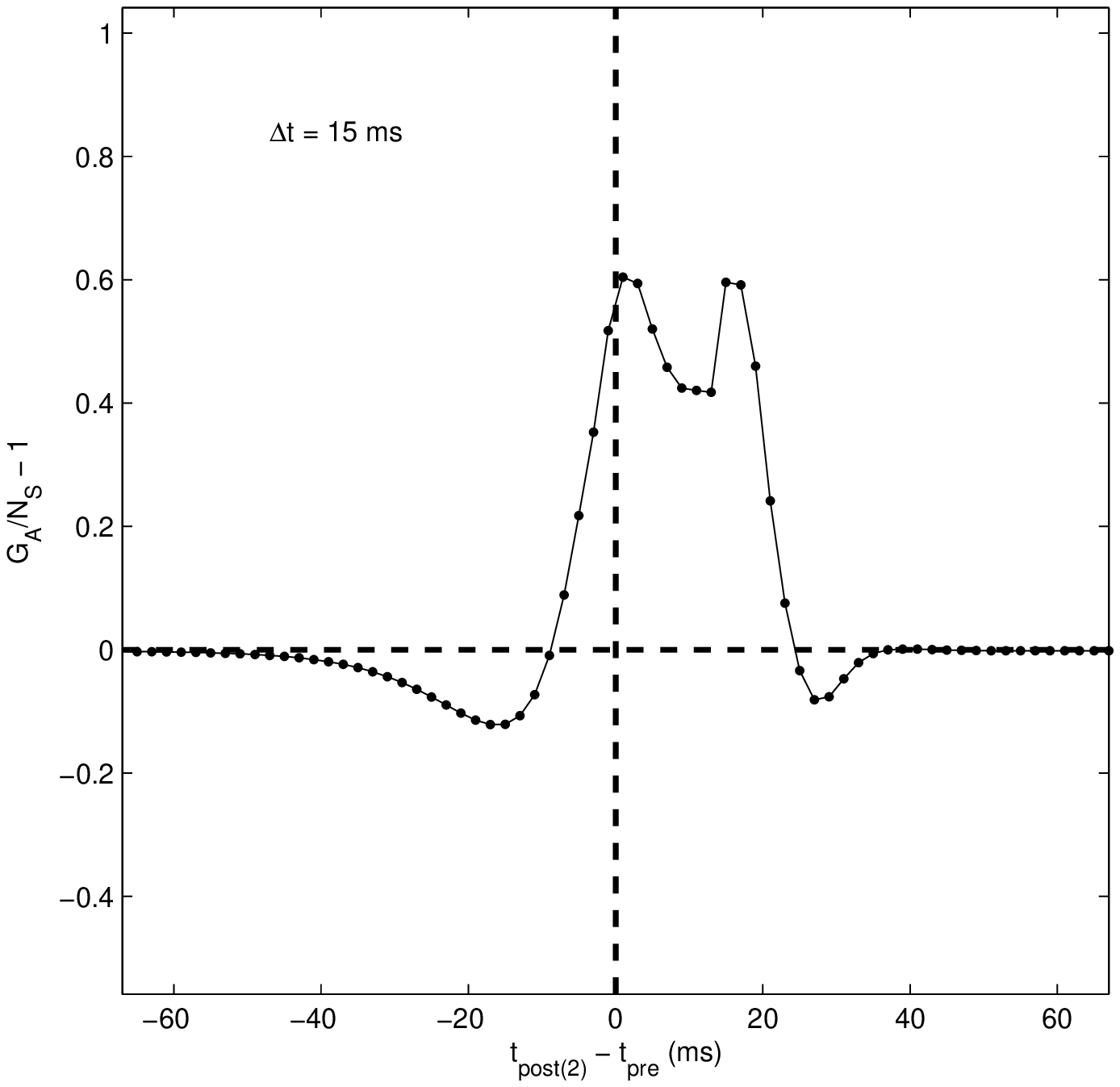}
\caption{ The change in normalized AMPA conductance per synapse,
$G_{AMPA}/N_{S}-1$, plotted as a function of the delay
$\tau=t_{post(2)}-t_{pre}$ (ms) when we present a single
presynaptic spike and a pair of postsynaptic spikes with a fixed time difference of 15 ms.
$t_{post(2)}$ represents the time of presentation of the second
postsynaptic spike. A distinct dip in the potentiated AMPA conductance is observed for times when the presynaptic 
spike falls in the time between the 2 postsynaptic spikes.
\label{fig 6}}
\end{figure}

\begin{figure}[ht!]
\includegraphics[width=3in,scale=.35,angle=0]{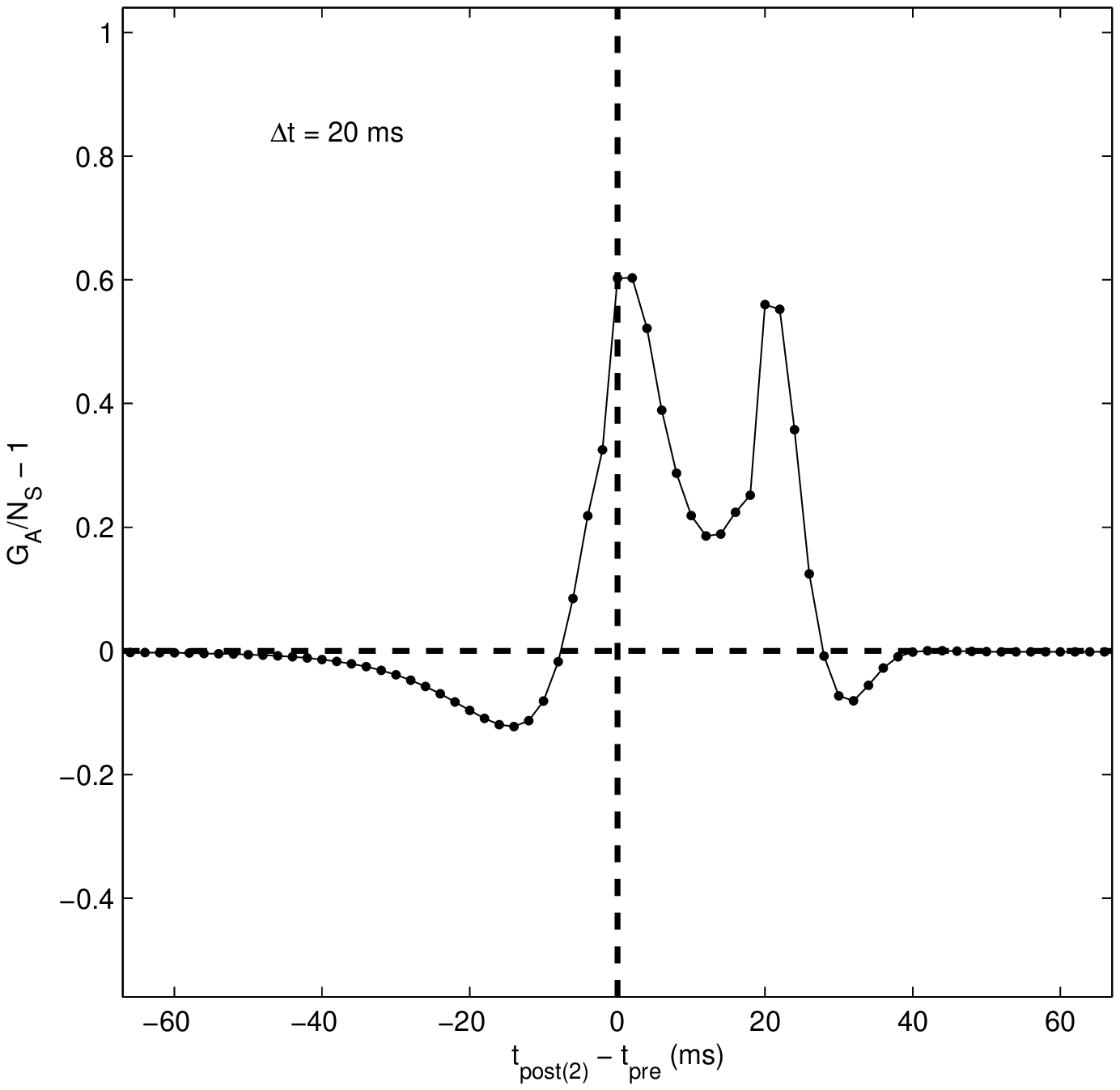}
\caption{ The change in normalized AMPA conductance per synapse,
$G_{AMPA}/N_{S}-1$, plotted as a function of the delay
$\tau=t_{post(2)}-t_{pre}$ (ms) when we select a single
presynaptic spike and a pair of postsynaptic spikes with a fixed time difference of 20 ms.
$t_{post(2)}$ represents the time of presentation of the second
postsynaptic spike. A distinct dip in the potentiated AMPA conductance is observed for times when the presynaptic
 spike falls in the time between the 2 postsynaptic spikes.
\label{fig 7}}
\end{figure}

\begin{figure}[ht!]
\includegraphics[width=3in,scale=.35,angle=-90]{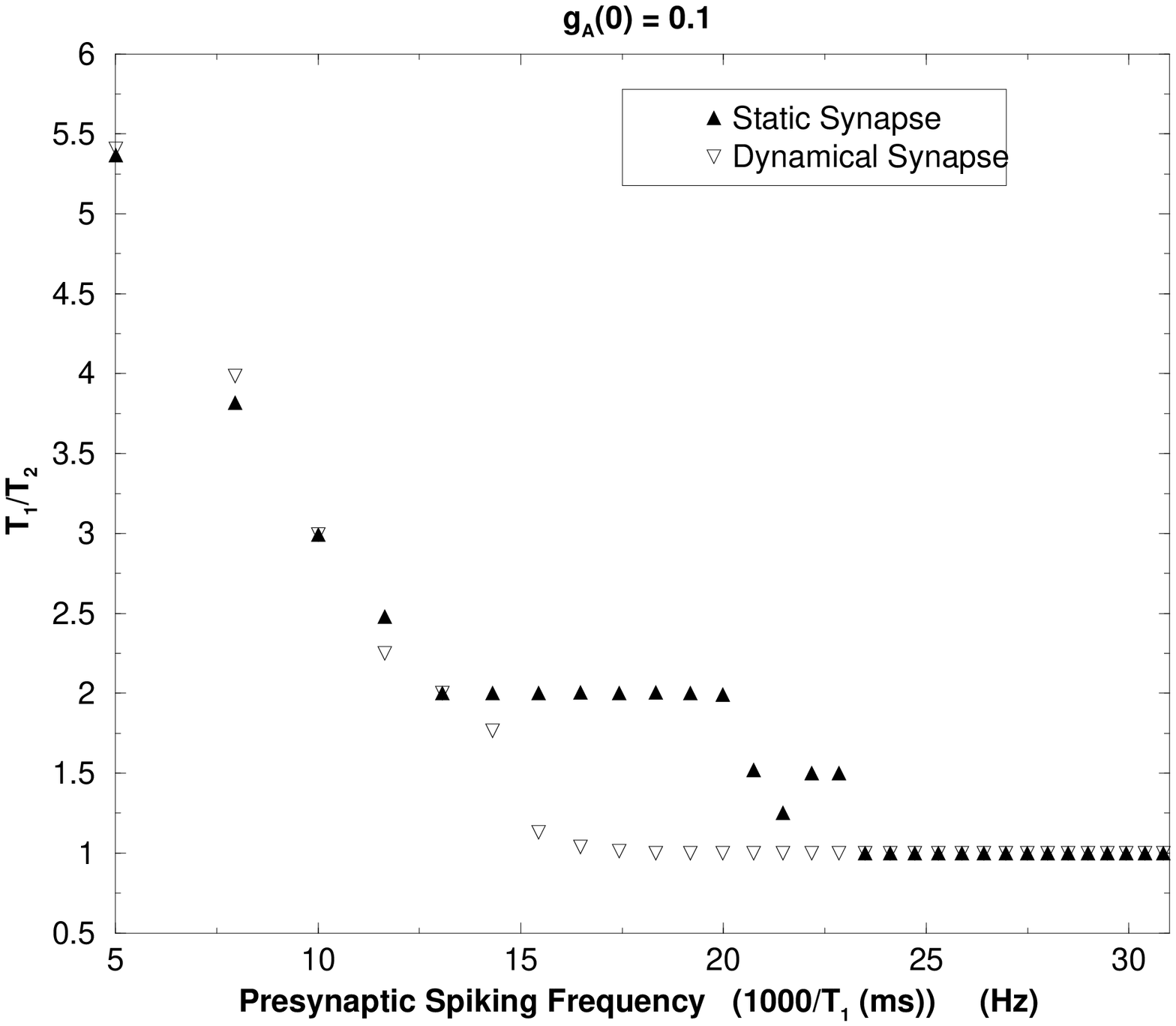}
\caption{ $T_{1}/T_{2}$, the ratio of the interspike interval $T_{1}$ of the presynaptic neuron to the 
interspike interval $T_{2}$ of the postsynaptic neuron, is plotted as a function of the
presynaptic input frequency, $1000/T_{1}$ Hz, for a synapse
starting at a base AMPA conductance of $g_{AMPA}(t=0)=0.1$
$mS/cm^{2}$. We see that the one-to-one synchronization window is
broadened when the static synapse is replaced by a plastic synapse
as determined by the three state model.
\label{fig 8}}
\end{figure}

\begin{figure}[ht!]
\includegraphics[width=3in,scale=.35,angle=-90]{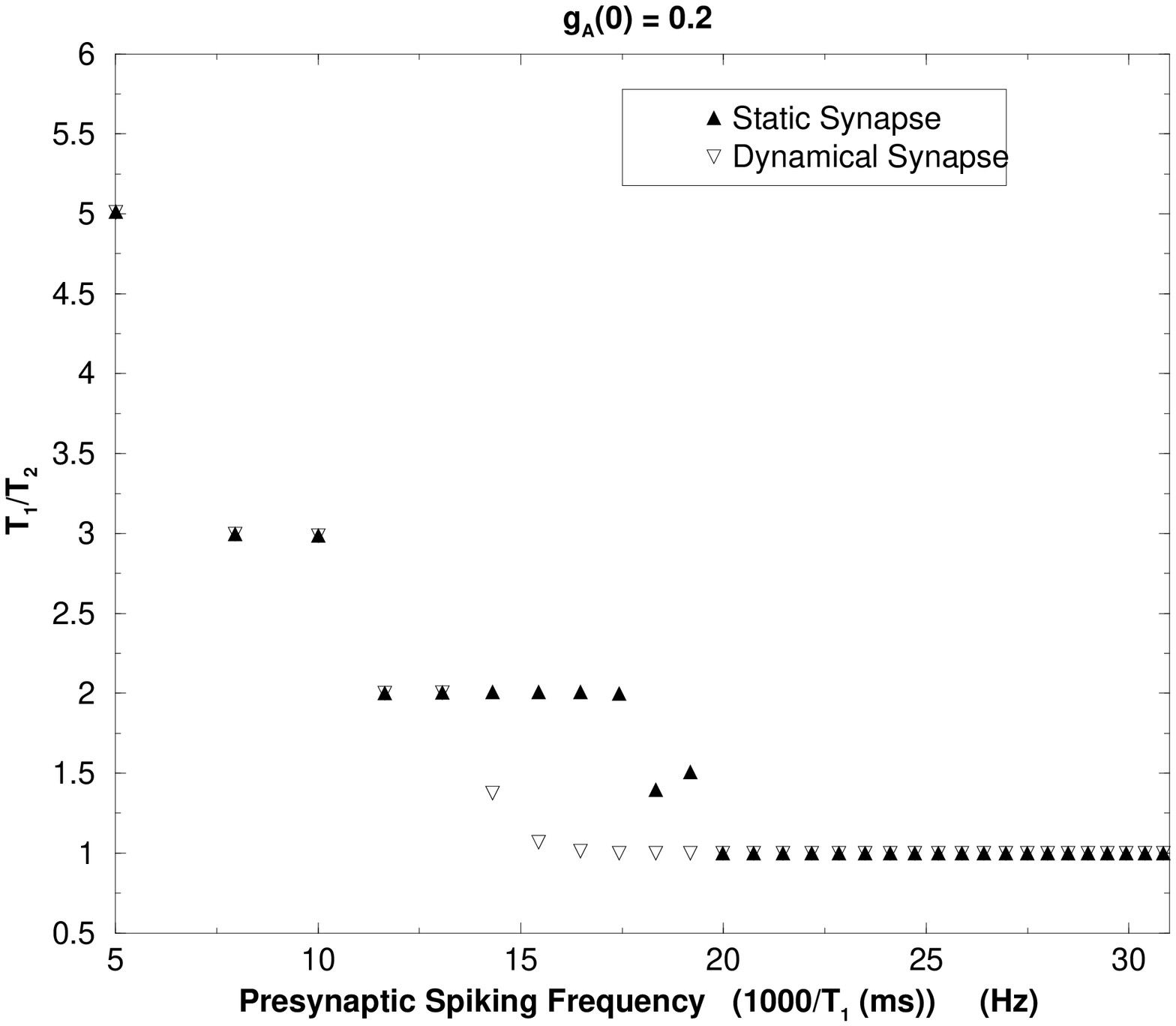}
\caption{ $T_{1}/T_{2}$, the ratio of the
interspike interval $T_{1}$ of the presynaptic neuron to the
interspike interval $T_{2}$ ms of the postsynaptic neuron, is
plotted as a function of the presynaptic input frequency,
$1000/T_{1}$ Hz, for a synapse starting at a base AMPA conductance
of $g_{AMPA}(t=0)=0.2$ $mS/cm^{2}$. We see that the one-to-one
synchronization window is broadened when the static synapse is
replaced by a plastic synapse as determined by the three state
model. 
\label{fig 9}}
\end{figure}

\begin{figure}[ht!]
\includegraphics[width=3in,scale=.35,angle=-90]{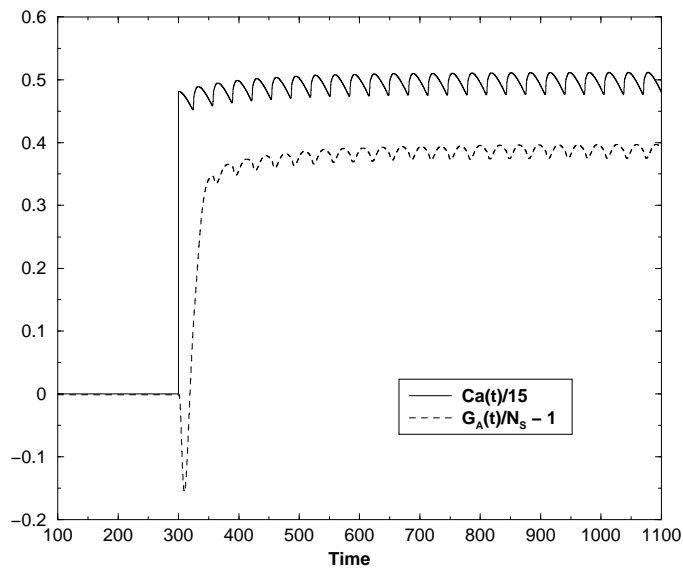}
\caption{Intracellular calcium concentration, scaled by 15, and the change in normalized
synaptic strength, $G_{AMPA}(t)/N_{S}-1$, is plotted as a function
of time in the case when a periodically spiking postsynaptic cell
is driven by a periodically spiking presynaptic input. 
\label{fig 10}}
\end{figure}

\begin{figure}[ht!]
\includegraphics[width=3in,scale=.35,angle=-90]{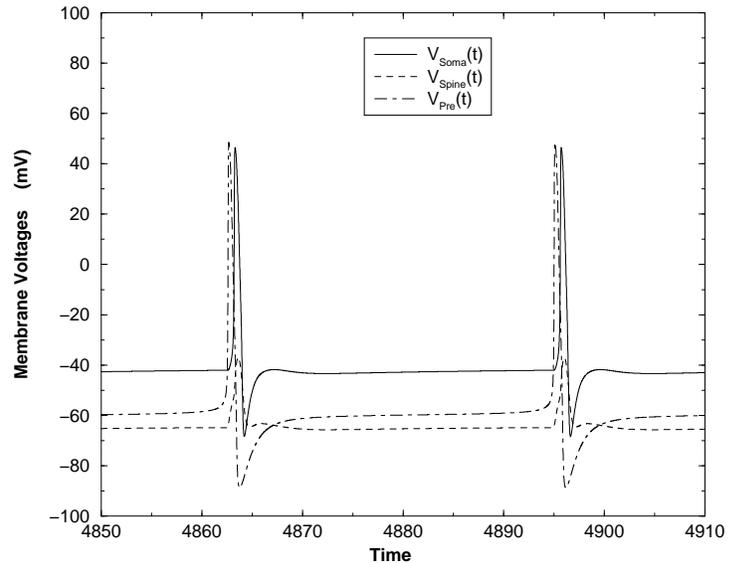}
\caption{$V_{\mbox{soma}}(t)$, $V_{\mbox{dendrite}}(t)$ and $V_{\mbox{pre}}(t)$, plotted as functions
of time, when the presynaptic and postsynaptic neurons are synchronized. Note that the presynaptic and 
postsynaptic neurons are synchronized in-phase with an internal, $V_{\mbox{soma}}(t)$ to $V_{\mbox{spine}}(t)$, 
time difference determined by the two compartments of the model neuron. 
\label{fig 11}}
\end{figure}

\end{document}